\documentclass[aps,prl,showpacs,twocolumn,superscriptaddress]{revtex4}
\usepackage{graphicx}
\usepackage{amsmath}
\usepackage{color}
\def\etal{$\it{et~al.}$}

\begin{document}

\title{The Nature of Quantum Hall States near the Charge Neutral Dirac Point in Graphene}

\author{Z. Jiang}
\altaffiliation{Electronic address: jiang@magnet.fsu.edu}
\affiliation{Department of Physics, Columbia University, New York, New York 10027, USA}
\affiliation{National High Magnetic Field Laboratory, Tallahassee, Florida 32310, USA}

\author{Y. Zhang}
\affiliation{Department of Physics, Columbia University, New York, New York 10027, USA}

\author{H. L. Stormer}
\affiliation{Department of Physics, Columbia University, New York, New York 10027, USA}
\affiliation{Department of Applied Physics and Applied Mathematics, Columbia University, New York, New York 10027, USA}
\affiliation{Bell Laboratories, Alcatel-Lucent, Murray Hill, New Jersey 07974, USA}

\author{P. Kim}
\affiliation{Department of Physics, Columbia University, New York, New York 10027, USA}

\date{\today}

\begin{abstract}
We investigate the quantum Hall (QH) states near the charge neutral Dirac point of a high mobility graphene sample in high magnetic fields. We find that the QH states at filling factors $\nu=\pm1$ depend only on the perpendicular component of the field with respect to the graphene plane, indicating them to be not spin-related. A non-linear magnetic field dependence of the activation energy gap at filling factor $\nu=1$ suggests a many-body origin. We therefore propose that the $\nu=0$ and $\pm1$ states arise from the lifting of the spin and sub-lattice degeneracy of the $n=0$ LL, respectively.
\end{abstract}

\maketitle

The experimental observation of the Quantum Hall (QH) effect in single atomic sheet of graphene \cite{novoselov05,zhang05} has attracted much attention recently, particularly due to the unique electronic transport observed in this material. Compared with the conventional integer QH effect in many other two-dimensional (2D) systems, the Hall resistance ($R_{xy}$) quantization condition in graphene is shifted by a half integer: $R_{xy}^{-1}=\pm g_s(n+1/2)\frac{e^2}{h}$, where $n$ is the Landau level (LL) index, $e$ is the electron charge, $h$ is Planck's constant, and $g_s=4$ is the LL degeneracy, accounting for spin and sub-lattice symmetry in graphene. This quantization condition leads to the QH effect appearing at filling factors $\nu=\pm2, \pm6, \pm10, \cdots$. It is now understood that this unique QH effect is related to the quasi-relativistic nature of the charge carriers in graphene \cite{zheng02,gusynin05-1,peres06}, stemming from the unusual linear dispersion relation of its bands near the charge neutral Dirac point in the graphene band structure \cite{wallace47,mcclure56}.

More recently, the QH effect in graphene has been studied in the extreme quantum limit in a very strong magnetic field \cite{zhang06}. New QH states, corresponding to filling factors $\nu=0, \pm1, \pm4$, are clearly resolved in magnetic fields $B>20$ T, indicating a lifting of the four-fold degeneracy of the $n=0$ LL and a two-fold degeneracy of the $n=\pm1$ LLs respectively. While angular dependent activation energy gap measurements indicate that the QH states at $\nu=\pm4$ are spin states, the origin of the QH states at $\nu=0, \pm1$ remains unresolved.

The nature of these QH states near the charge neutral Dirac point is of fundamental interest. There have been numerous theoretical investigations \cite{khveshchenko01,gusynin05-2,kane05,neto06,abanin06,nomura06,alicea06,yang06,goerbig06,herbut06,gusynin06,ezawa06,yang07,fertig06,apalkov06,fuchs07,lukose06} of these states, and their origin is currently under considerable debate. Recently, Abanin \etal $\ $\cite{abanin07} suggested that the $\nu=0$ QH state is spin-polarized and dissipative, owing to counter-propagating edge states at the charge neutral point, supported by a finite metallic resistivity at low temperatures.

In this Letter, we present an experimental investigation of the QH states near the Dirac point. We find that the $\nu=\pm1$ states depend only on the out of plane component of the applied magnetic field, and show a rather large energy gap with an approximately square root dependence on the magnetic field. This suggests a many-particle origin of this splitting as it would originate from the breaking of the sub-lattice degeneracy of the $n=0$ LL at the Dirac point. As a consequence, and by elimination, our results would imply that the $\nu=0$ QH gap is induced by the lifting of the spin degeneracy.

Our sample is a high quality graphene specimen with mobility as high as $\sim2\times10^4$ cm$^2$/Vs measured at carrier density $n_e=4\times10^{12}$ cm$^{-2}$. The graphene sheet is mechanically extracted from Kish graphite following a method similar to the one described in \cite{novoselov05,zhang05}. The sample is deposited onto a Si substrate, which serves as a gate electrode separated from the sample by 300 nm of insulating SiO$_2$. To perform transport measurement, multiple electrodes are patterned in van der Pauw geometry (inset to Fig. 1) using conventional electron beam lithography, followed by Au/Cr (30/3 nm) thermal evaporation and a standard lift-off process. The electronic transport is measured over the temperature range of 4.2-300 K, using a lock-in technique. The sample is mounted on a single-axis tilting stage to allow \textit{in situ} tuning of the angle, $\theta=\text{cos}^{-1}(B_p/B_{tot})$, where $B_{tot}$ is the total magnetic field and $B_p$ is the component perpendicular to the graphene plane.

The four-fold degeneracy of the $n=0$ LL of graphene consists of a two-fold degeneracy from the spin symmetry and a two-fold degeneracy from the sub-lattice symmetry. One may be able to distinguish the origin of any particular splitting by performing magnetotransport measurement in a tilted field, where a spin splitting depends on $B_{tot}$, whereas a sub-lattice splitting (caused by electron-electron correlations) would only depend on $B_p$. Figure 1 shows the measured magnetoresistance, $R_{xx}$, with respect to the back gate voltage, $V_g$, at a temperature of $T=4.2$ K and in two different magnetic fields: $B_{tot}=45$ T, $B_p=20$ T (dashed curve); and $B_{tot}=30$ T, $B_p=20$ T (solid curve). Note that in this comparison we use the same $B_p$ but different $B_{tot}$. The minimum magnetoresistance $R_{xx}^{min}$ substantially increases as $B_{tot}$ decreases for $\nu=-4$ state, while for the $\nu=\pm1$ states $R_{xx}^{min}$ hardly varies with $B_{tot}$ for the same $B_p$. This observation supports the spin related origin of the $\nu=\pm4$ splittings \cite{zhang06}, and also, more importantly, suggests that the $\nu=\pm1$ states are likely due to the breaking of the orbital degeneracy of the sub-lattice symmetry in the $n=0$ LL.

\begin{figure}[t]
\includegraphics[width=8.5cm]{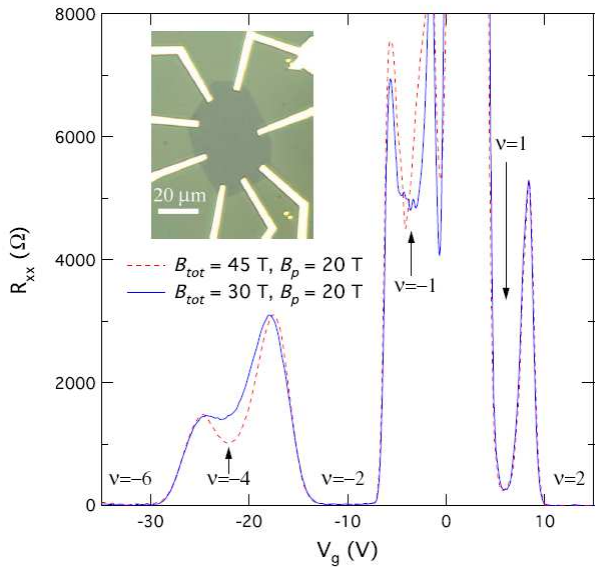}
\caption{(color online). Magnetoresistance, $R_{xx}$, with respect to the back gate voltage, $V_g$, at a temperature of $T=4.2$ K and in two different magnetic fields: $B_{tot}=45$ T, $B_p=20$ T (dashed curve); and $B_{tot}=30$ T, $B_p=20$ T (solid curve). The minimum magnetoresistance $R_{xx}^{min}$ substantially increases as $B_{tot}$ decreases for $\nu=-4$ state, while for the $\nu=\pm1$ states $R_{xx}^{min}$ practically does not depend on $B_{tot}$ for the same $B_p$.}
\end{figure}
	
In order to further characterize the nature of $\nu=\pm1$ QH states, we measure the activation energy of $R_{xx}^{min}$ at fixed magnetic fields. Figure 2(b) displays the Arrhenius plots of $R_{xx}^{min}$ of the $\nu=1$ state \cite{note1}. A well-defined thermal activation behavior is readily observable ($R_{xx}^{min}\sim\text{exp}[-\Delta E/2k_BT]$, where $k_B$ is the Boltzmann constant), and the corresponding energy gap, $\Delta E$, can be extracted for different magnetic fields. In Fig. 2(a), we plot the obtained energy gaps at $\nu=1$, denoted as $\Delta E(\nu=1)$, as a function of $B$-field. For comparison, we have also reproduced the measured $\Delta E(\nu=\pm4)$ of the same sample from Ref. \cite{zhang06}. We find $\Delta E(\nu=1)$ to be considerably larger than $\Delta E(\nu=\pm4)$. For instance at 45 T, $\Delta E(\nu=1)\geq4\Delta E(\nu=\pm4)$. Moreover, unlike for $\Delta E(\nu=\pm4)$, which showed a linear $B$-field dependence, the $B$-field dependence of $\Delta E(\nu=1)$ does not seem to follow such a simple relationship. Forcing a linear fit onto $\Delta E(\nu=1)$ produces a positive $y$-intercept which would indicate an unphysical, negative LL energy width. In fact, a $\sqrt{B}$ behavior provides a better fit to the $\Delta E(\nu=1)$ data as shown in the solid curve in Fig. 2(a).

\begin{figure}[t]
\includegraphics[width=8.5cm]{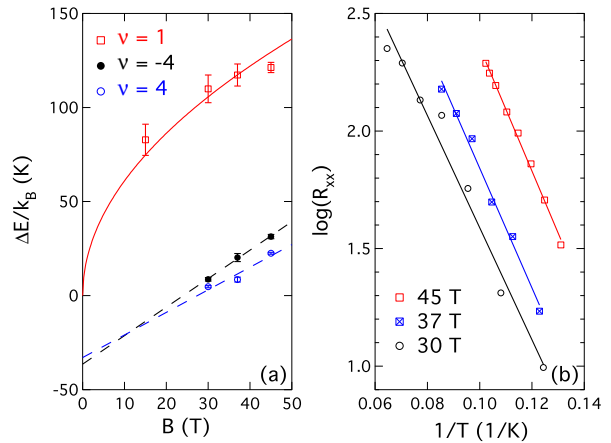}
\caption{(color online). (a) The measured activation energy gap, $\Delta E$, as a function of magnetic field for the QH states at filling factors $\nu=1$ (square), $\nu=-4$ (solid dot), and $\nu=4$ (open dot). While $\Delta E(\nu=\pm4)$ are linear with respect to $B$-field (data are reproduced from Ref. \cite{zhang06}), $\Delta E(\nu=1)$ can be fitted better by a $\sqrt{B}$ dependence (solid curve). (b) Arrhenius plots of $R_{xx}$ minimum of the $\nu=1$ state for three different magnetic fields: 45 T, 37 T, and 30 T. The straight lines are linear fits to the data.}
\end{figure}

The lack of a linear dependence and the existence of a roughly $\sqrt{B}$ dependence point to a non-spin origin and possibly to a many-particle origin of the gap, and suggest that the $\nu=\pm1$ states are associated with a spontaneous breaking of the sub-lattice symmetry driven by the electron-electron interactions \cite{abanin06,nomura06,alicea06,yang06,goerbig06,herbut06,gusynin06,ezawa06,yang07}. In this picture, $\Delta E(\nu=\pm1)$ is expected to be on the scale of $e^2/\epsilon l_B$ \cite{yang06}, and thus proportional to $\sqrt{B}$, where $l_B=\sqrt{\hbar/eB}$ is the magnetic length and $\epsilon$ is the dielectric constant. We calculate that $e^2/\epsilon l_B\sim1100$ K, assuming $\epsilon=4$ and $B=45$ T. This value is much larger than the Zeeman energy $E_Z=g\mu_BB\sim60$ K at $B=45$ T, where $g=2$ is the $g$-factor and $\mu_B$ is the Bohr magneton. This simple evaluation indicates that $\Delta E(\nu=\pm1)>>E_Z$ in the experimentally accessible magnetic field, suggesting the importance of electron-electron interaction under mangetic fields.

In Fig. 3, we summarize our current understanding of the sequence of the QH states near the charge neutral Dirac point of graphene in a schematic of the LL hierarchy. We use up-arrows and down-arrows to represent the spin of the charge carries, and solid (blue) and open (red) dots for different valleys in the graphene band structure. Since our measurements suggest that the $\nu=\pm1$ states are associated with the valley splitting of the $n=0$ LL due to electron-electron correlations, the QH state at $\nu=0$ must be related to the spin splitting of this LL. However, we also note that the behavior of $R_{xx}$ and $R_{xy}$ at $\nu=0$ is completely different from that of any other QH states away from the charge neutral Dirac point. Unlike usual QH states, the $\nu=0$ QH state does not show a resistance minimum in $R_{xx}$ nor a clear resistance plateau in $R_{xy}$. This state only becomes visible as a plateau in the Hall conductance. Figure 4 displays $R_{xx}$ vs. $V_g$ near the Dirac point over a wide range of temperatures. No activation behavior has been observed at $\nu=0$. Recently, Abanin \etal $\ $provided a possible interpretation for the existence of this state as the consequence of counter-circulating edge states with opposite spin \cite{abanin07}. Such a state would be consistent with our proposal shown in Fig. 3.

\begin{figure}[t]
\includegraphics[width=8cm]{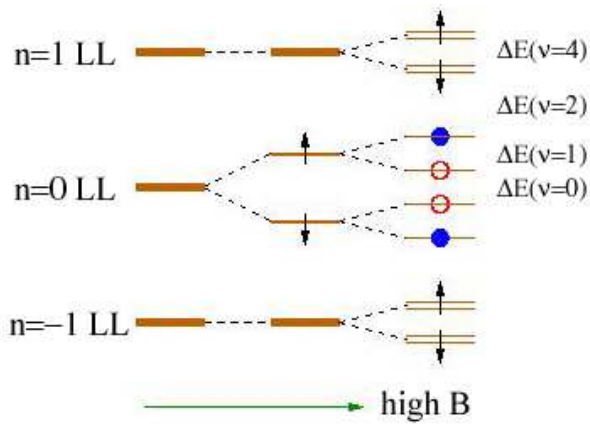}
\caption{(color online). Schematic of the LL hierarchy in graphene in magnetic fields. The up and down arrows represent the spin of the charge carries, and the solid (blue) and open (red) dots indicate different valleys in the graphene electronic band.}
\end{figure}

We now address the relative size of the energy gap between the levels displayed in Fig. 3. In a sufficiently large magnetic field, typically $B>20$ T, the QH states in high-quality graphene specimens are robust, even at room temperature \cite{novoselov07} (see also Fig. 4). From the inset to Fig. 4, we estimate the activation energy gaps of the QH states at $\nu=\pm2$ and $B=45$ T to be $\Delta E(\nu=2)\approx890$ K (solid dots), and $\Delta E(\nu=-2)\approx570$ K (open dots). In a simplistic view, we may interpret these values as the gap between the $n=0$ and the $n=\pm1$ LLs in Fig. 3. In a single particle picture with un-lifted spin and sub-lattice degeneracies \cite{haldane88}, the LL energy spectrum in graphene can be described by 
\begin{equation}
E_n=\text{sgn}(n)\sqrt{2e\hbar v_F^2B\left|n\right|},
\end{equation}
where $v_F$ is the Fermi velocity of graphene with a typical value of $v_F\approx10^6$ m/s \cite{novoselov05,zhang05,dresselhaus02,brandt88}. Hence, at $B=45$ T, the calculated energy spacing between the $n=0$ and the $n=1$ LL would be $E_{0\rightarrow1}\approx2800$ K. This value is more than three times larger than the largest measured energy gap $\Delta E(\nu=2)\approx890$ K. We believe that such a large discrepancy cannot be accounted for in terms of spin or sub-lattice symmetry splitting, nor by any reasonable LL broadening. We first rule out the possibility of an enhanced spin splitting since the $g$-factor is not enhanced by exchange in a completely filled LL. With the bare value of $g=2$, the spin splitting of the LL reaches only $g\mu_BB\sim60$ K at 45 T. We further eliminate the sub-lattice degeneracy splitting as a potential explanation of the observed discrepancy, as this gap $\Delta E(\nu=\pm1)$ must collapse when the Fermi energy lies between the $n=0$ and the $n=1$ LL. Finally, the LL broadening due to scattering may lead to a reduction of the energy gap. However, we infer such a reduction is negligible compared to $E_{0\rightarrow1}\approx2800$ K: from our low temperature measurements ($T=1.4$ K) \cite{zhang06}, we estimate a LL broadening of $\Gamma\approx20$ K. Since the mobility of graphene changes only by $\sim30\%$ from 30 mK to room temperature \cite{tan}, $\Gamma$ is likely irrelevant on the scale of $E_{0\rightarrow1}$.

\begin{figure}[t]
\includegraphics[width=8.5cm]{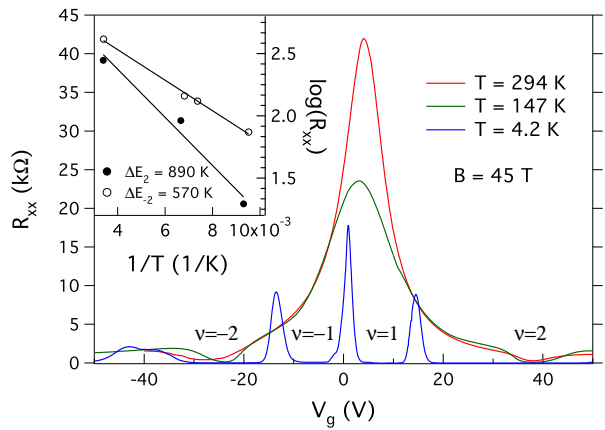}
\caption{(color online). Magnetoresistance, $R_{xx}$, as a function of the back gate voltage, $V_g$, at $B=45$ T in a wide temperature range, from room temperature to liquid helium temperature. A robust QH effect is clearly seen even at room temperature. No activated behavior has been observed in $R_{xx}$ of the $\nu=0$ state (at Dirac point). Inset: Arrhenius plots of $R_{xx}$ minimum of the $\nu=\pm2$ states at $B=45$ T. The straight lines are linear fits to the data.}
\end{figure}

At this stage we do not know what causes this strong reduction of the measured LL energy gap as compared to the calculated one. One may speculate that many-particle effects are partially responsible for the discrepancy. Recent infrared experiments in graphene \cite{sadowski06,jiang07,deacon07} between LL levels $n=0$ and $n=\pm1$ yield a rather good agreement with Eq. 1. Yet theory suggests that as much as 30$\%$ of this energy is due to many-particle corrections \cite{iyengar07}. Since both single-particle energy level and many-particle correction have $\sqrt{B}$-field dependences in graphene, they cannot be separated. This would suggest that the Fermi velocity (typically $v_F\approx10^6$ m/s) can be as much as $\sim30\%$ smaller in reality, the difference being made up for by many-particle corrections. If activation energy measurements were much less affected by such corrections, the expected gaps would follow the bare value of $v_F$, leading to a $\sim30\%$ reduction from Eq. 1. This would bring the calculated energy gap of $E_{0\rightarrow1}\approx1900$ K closer to the measured gap of $\Delta E(\nu=2)\approx890$ K, but still leaves a substantial, unresolved discrepancy. The implication of such a reinterpretation of $v_F$ would be considerable and more extensive studies of the LL spectrum in graphene will be required to verify such a trend.

In conclusion, we study the QH states in graphene at filling factors $\nu=\pm1$ in tilted magnetic fields and elevated temperatures. Our results indicate that the $\nu=\pm1$ QH states originate from the lifting of the sub-lattice symmetry of the $n=0$ LL caused by electron-electron interactions. Measurements of the activation energy gaps of the QH states near the Dirac point indicate a significant deviation from a simplistic single-particle model, which suggests that many-particle effects need to be taken into account to understand the LLs near the charge neutral Dirac point.

We would like to thank I. Aleiner, K. Yang, A.K. Geim, and K. Nomura for useful discussions. This work is supported by the DOE (DE-AIO2-04ER46133 and DE-FG02-05ER46215), NSF (DMR-03-52738 and CHE-0117752), ONR (N000150610138), NYSTAR, the Keck Foundation, and the Microsoft Project Q. A portion of this work was performed at the National High Magnetic Field Laboratory, which is supported by NSF Cooperative Agreement No. DMR-0084173, by the State of Florida, and by the DOE. We thank L.W. Engel, S.T. Hannahs, E.C. Palm, T.P. Murphy, G.E. Jones, J. Jaroszynski, and B.L. Brandt for experimental assistance.


\begin{thebibliography}{text}
\bibitem{note1}We find that the QH state at $\nu=-1$ is not fully developed compared to the state at $\nu=1$. In particular, we find that $R_{xx}^{min}\sim80$ $\Omega$ at $\nu=-1$, while $R_{xx}^{min}<1$ $\Omega$ at $\nu=1$. This asymmetry may be due to the charge disorders between the graphene sheet and the SiO$_2$/Si substrate, which may have different scattering rates for electron/hole carriers in graphene.
\end{thebibliography}
\end{document}